\PassOptionsToPackage{unicode}{hyperref}
\PassOptionsToPackage{hyphens}{url}
\PassOptionsToPackage{dvipsnames,svgnames,x11names}{xcolor}
\documentclass[
  number,
  preprint,
  5p,
  twocolumn]{elsarticle}

\usepackage{amsmath,amssymb}
\usepackage{iftex}
\ifPDFTeX
  \usepackage[T1]{fontenc}
  \usepackage[utf8]{inputenc}
  \usepackage{textcomp} 
\else 
  \usepackage{unicode-math}
  \defaultfontfeatures{Scale=MatchLowercase}
  \defaultfontfeatures[\rmfamily]{Ligatures=TeX,Scale=1}
\fi
\usepackage{lmodern}
\ifPDFTeX\else  
\fi
\IfFileExists{upquote.sty}{\usepackage{upquote}}{}
\IfFileExists{microtype.sty}{
  \usepackage[]{microtype}
  \UseMicrotypeSet[protrusion]{basicmath} 
}{}
\makeatletter
\@ifundefined{KOMAClassName}{
  \IfFileExists{parskip.sty}{%
    \usepackage{parskip}
  }{
    \setlength{\parindent}{0pt}
    \setlength{\parskip}{6pt plus 2pt minus 1pt}}
}{
  \KOMAoptions{parskip=half}}
\makeatother
\usepackage{xcolor}
\makeatletter
\ifx\paragraph\undefined\else
  \let\oldparagraph\paragraph
  \renewcommand{\paragraph}{
    \@ifstar
      \xxxParagraphStar
      \xxxParagraphNoStar
  }
  \newcommand{\xxxParagraphStar}[1]{\oldparagraph*{#1}\mbox{}}
  \newcommand{\xxxParagraphNoStar}[1]{\oldparagraph{#1}\mbox{}}
\fi
\ifx\subparagraph\undefined\else
  \let\oldsubparagraph\subparagraph
  \renewcommand{\subparagraph}{
    \@ifstar
      \xxxSubParagraphStar
      \xxxSubParagraphNoStar
  }
  \newcommand{\xxxSubParagraphStar}[1]{\oldsubparagraph*{#1}\mbox{}}
  \newcommand{\xxxSubParagraphNoStar}[1]{\oldsubparagraph{#1}\mbox{}}
\fi
\makeatother

\usepackage{longtable,booktabs,array}
\usepackage{calc} 
\usepackage{etoolbox}
\makeatletter
\patchcmd\longtable{\par}{\if@noskipsec\mbox{}\fi\par}{}{}
\makeatother
\IfFileExists{footnotehyper.sty}{\usepackage{footnotehyper}}{\usepackage{footnote}}
\makesavenoteenv{longtable}
\usepackage{graphicx}
\makeatletter
\def\maxwidth{\ifdim\Gin@nat@width>\linewidth\linewidth\else\Gin@nat@width\fi}
\def\maxheight{\ifdim\Gin@nat@height>\textheight\textheight\else\Gin@nat@height\fi}
\makeatother
\setkeys{Gin}{width=\maxwidth,height=\maxheight,keepaspectratio}
\makeatletter
\def\fps@figure{htbp}
\makeatother

\usepackage{fancyhdr}
\newcommand{\thetool}{FBCV}
\makeatletter
\@ifpackageloaded{caption}{}{\usepackage{caption}}
\AtBeginDocument{%
\ifdefined\contentsname
  \renewcommand*\contentsname{Table of contents}
\else
  \newcommand\contentsname{Table of contents}
\fi
\ifdefined\listfigurename
  \renewcommand*\listfigurename{List of Figures}
\else
  \newcommand\listfigurename{List of Figures}
\fi
\ifdefined\listtablename
  \renewcommand*\listtablename{List of Tables}
\else
  \newcommand\listtablename{List of Tables}
\fi
\ifdefined\figurename
  \renewcommand*\figurename{Figure}
\else
  \newcommand\figurename{Figure}
\fi
\ifdefined\tablename
  \renewcommand*\tablename{Table}
\else
  \newcommand\tablename{Table}
\fi
}
\@ifpackageloaded{float}{}{\usepackage{float}}
\floatstyle{ruled}
\@ifundefined{c@chapter}{\newfloat{codelisting}{h}{lop}}{\newfloat{codelisting}{h}{lop}[chapter]}
\floatname{codelisting}{Listing}

\makeatother
\makeatletter
\@ifpackageloaded{caption}{}{\usepackage{caption}}
\@ifpackageloaded{subcaption}{}{\usepackage{subcaption}}
\makeatother
\usepackage{float}
\makeatletter
\let\oldlt\longtable
\let\endoldlt\endlongtable
\def\longtable{\@ifnextchar[\longtable@i \longtable@ii}
\def\longtable@i[#1]{\begin{figure}[H]
\onecolumn
\begin{minipage}{0.5\textwidth}
\oldlt[#1]
}
\def\longtable@ii{\begin{figure}[H]
\onecolumn
\begin{minipage}{0.5\textwidth}
\oldlt
}
\def\endlongtable{\endoldlt
\end{minipage}
\twocolumn
\end{figure}}
\makeatother
\journal{ASA Student Paper Competition}

\ifLuaTeX
  \usepackage{selnolig}  
\fi
\usepackage[]{natbib}
\usepackage{bookmark}

\IfFileExists{xurl.sty}{\usepackage{xurl}}{} 
\hypersetup{
  pdftitle={Interactive Visualization Framework for Forensic Bullet Comparisons},
  pdfauthor={Nathan Rethwisch; Heike Hofmann},
  pdfkeywords={data visualization, interactive forensic
modeling, cross-correlation function, land engraved area, forensic
pattern analysis, forensic statistics},
  colorlinks=true,
  linkcolor={blue},
  filecolor={Maroon},
  citecolor={Blue},
  urlcolor={Blue},
  pdfcreator={LaTeX via pandoc}}

\begin{document}

\begin{frontmatter}
\title{Interactive Visualization Framework for Forensic Bullet
Comparisons}
\author[1]{Nathan Rethwisch%
\corref{cor1}%
}
 \ead{nreth@iastate.edu} 
\author[2]{Heike Hofmann%
}
 \ead{hhofmann4@unl.edu} 

\affiliation[1]{organization={Iowa State University, Department of
Statistics},addressline={2438 Osborn Dr, Ames IA, 50010},postcodesep={}}
\affiliation[2]{organization={University of Nebraska Lincoln, Department
of Statistics},addressline={3310 Holdrege St, Lincoln, NE,
68583},postcodesep={}}

\cortext[cor1]{Corresponding author}

\begin{abstract}
The current method for forensic analysis of bullet comparison relies on
manual examination by forensic examiners to determine if bullets were
discharged from the same firearm. This process is highly subjective,
prompting the development of algorithmic methods to provide objective
statistical support for comparisons. However, a gap exists between the
technical understanding of these algorithms and the typical background
of many forensic examiners. We present a visualization tool designed to
bridge this gap, allowing for the presentation of statistical
information in a more familiar format to forensic professionals. The
forensic bullet comparison visualizer (FBCV) features a variety of plots
that will enable the user to examine every step of the algorithmic
comparison process. We demonstrate the utility of the FBCV by applying
it to data from the Houston Science Lab, where it helped identify an
error in the comparison process caused by mislabeling. This tool can be
used for future investigations, such as examining how distance between
shots affects scores. The FBCV offers a user-friendly way to convey
complex statistical information to forensic examiners, facilitating
their understanding and utilization of algorithmic comparison methods.
\end{abstract}

\end{frontmatter}

\clearpage
\newpage

\setcounter{page}{1}
\pagestyle{fancy}
\fancyhead{} 
\fancyfoot{} 
\fancyhead[L]{\sc{Interactive Visualization Framework for Forensic Bullet Comparisons}}
\fancyfoot[R]{\thepage}

\section{Introduction and Background}\label{introduction-and-background}

Identifying the firearm used in a crime is a critical component of
forensic investigations and plays a pivotal role in criminal
investigations. Ensuring evidence is appropriately identified is crucial
in upholding the integrity of the criminal justice system.

Current forensic practices rely on examiners to visually inspect bullets
under a comparison microscope for similarities of marks on the bullets'
surfaces. As a bullet is discharged from the firearm, the rifling in the
barrel forces the bullet to follow the groove pattern like rails.
Micro-imperfections in the barrel leave scratches (called
\emph{striations}) on the bullet's surface. Striations on land engraved
areas (LEA; the area between two grooves) are assumed to be unique to
the individual firearm. This allows forensic examiners to determine
whether two bullets originate from the same source by seeing if these
LEAs match \citep{afte}. However, this process is highly subjective,
relying heavily on an examiner's expertise \citep{nas2009, pcast}. These
criticisms triggered the development of algorithmic comparisons
\citep{carriquiryMachineLearningForensic2019, chenFiredBulletSignature2019, chuAutomaticIdentificationBullet2013, juOpenSourceImplementationCMPS2022, vorburgerApplicationsCrosscorrelationFunctions2011, vorburgerTopographyMeasurementsApplications2015}
to provide objective measures with the goal of augmenting an examiner's
testimony.

Algorithmic comparison methods have demonstrated considerable potential
to quantify the similarity between pairwise pieces of evidence. However,
current approaches have created a gap between the statistical metrics
and the practical understanding of these metrics by practitioners. This
gap highlights the need for a more effective method to assist forensic
practitioners in assessing and understanding the algorithm's
performance. Here, we are proposing an interactive interface designed to
visualize the statistical metrics embedded in the context of the data
\citep{wickhamVisualizingStatisticalModels2015} in a manner that is
intuitive and accessible to forensic examiners. The forensic bullet
comparison visualizer (FBCV) combines a set of interactive
visualizations, allowing forensic examiners to engage with the complex
algorithmic data at each stage of the process, thereby bridging the gap
between statistical analysis and practical forensic application.

This paper presents a short review of the algorithmic comparison
process. We then discuss the choice of visuals in supporting diagnostics
at each stage of the process. We also showcase the diagnostic
capabilities of these visuals by presenting a real-world use case where
we successfully applied the FBCV to identify an error in the
data-cleaning process.

The data used for illustrating the process is a dataset of scans
provided by a collaboration of CSAFE (Center for Statistics and
Applications of Forensic Evidence) and the Houston Forensic Science
Center (HFSC). The data consists of scans from 40 test fires of each of
13 Ruger LCP barrels. Ten barrels (labeled `A' through `J') were
consecutively manufactured, while the remaining three (labeled 1-3) come
from HFSC's reference library of firearms. Here, we are analyzing 40
sequential shots from each of the barrel. The lettered barrels were only
fired ten times each before this study. For ease of notation, we refer
to these shots as 11 through 50. LCP barrels are traditionally rifled
barrels with 6 lands and 6 groove areas. These barrels mark well, i.e.,
striation marks are almost visible to the naked eye, making them
well-suited for a forensic analysis. Scans of the bullets were obtained
using a high-resolution confocal light microscope. For each bullet, 3d
topographic images of each of the six land engraved areas (LEA) were
acquired at 20x magnification (corresponding to 0.645 micron/pixel),
resulting in a total of 3,120 LEA scans (13 barrels x 40 bullets x 6
lands).

These scans provide the basis for algorithmic comparisons. For the
processing of scans and comparison of signals, we follow the steps
outlined in \citep{hareAutomaticMatchingBullet2017}, implemented in the
\texttt{bulletxtrctr} package in R
\citep{hofmannBulletxtrctrAutomaticMatching2022}. We apply the following
steps to each of the 3,120 scans {[}steps 1-4{]} and each pair of scans
{[}step 5{]}. The results from these comparisons were then rendered in
our visualization framework Section~\ref{sec-visuals}.

\begin{figure*}

\centering{

\includegraphics[width=6.39in,height=\textheight]{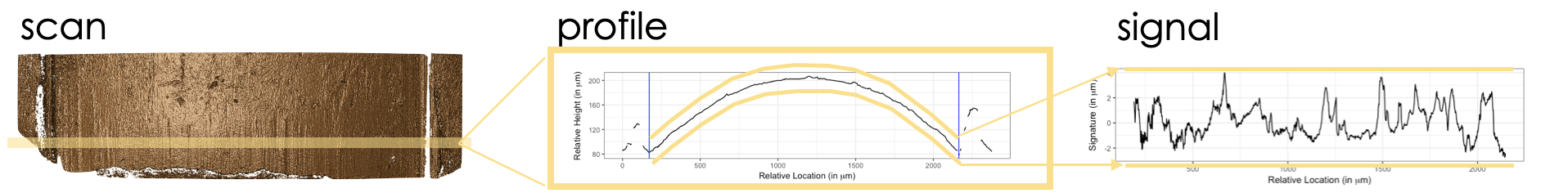}

}

\caption{\label{fig-extracted-sigs}Signal extraction from a 3d
topographic scan. From left to right we see a rendering of a 3d
topographic scan of a land-engraved area, the profile corresponding to
the horizontal yellow line, and the signal resulting from removing the
bullet curvature from the profile.}

\end{figure*}%

\begin{figure}

\centering{

\includegraphics[width=0.33\textwidth,height=\textheight]{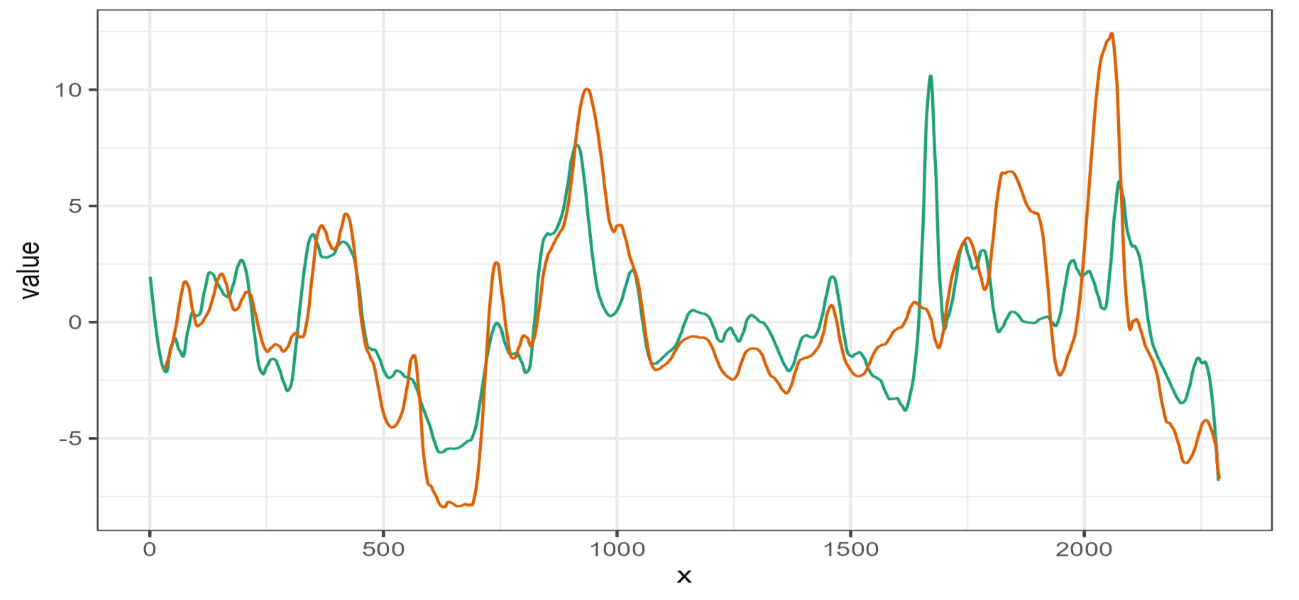}

}

\caption{\label{fig-aligned-sigs}Aligned Signals from lands of two
separate bullets.}

\end{figure}%

\begin{enumerate}
\def\labelenumi{\arabic{enumi}.}
\item
  A 3d LEA scan (left in Figure~\ref{fig-extracted-sigs}) is inspected
  for its suitability for comparisons, scans of low quality or of
  damaged lands (due to tank `rash', pitting, or cracks) are removed
  from the analysis.
\item
  A crosscut is chosen orthogonal to the direction of well-marked
  striations (marked as a yellow line in the rendering of the scan left
  in Figure~\ref{fig-extracted-sigs}).
\item
  The topographical measurements corresponding to this crosscut are
  extracted from the scan. The middle plot in
  Figure~\ref{fig-extracted-sigs} shows the profile of these height
  measurements along the crosscut. The spike in height at either end of
  the profile indicates the start of the neighboring groove areas and
  need to be excluded from the comparison (values outside the vertical
  blue lines).
\item
  The signal for a LEA (shown on the right of
  Figure~\ref{fig-extracted-sigs}) is created by removing the bullet
  curvature from the profile using a non-parametric smooth
  \citep{clevelandRobustLocallyWeighted1979}.
\item
  Finally, signals are aligned pairwise, as shown in
  Figure~\ref{fig-aligned-sigs}, and metrics assessing their similarity
  --such as the number of matching peaks, height of matching peaks,
  number of consecutively matching peaks, and, more statistically,
  cross-correlation-- are extracted.
\end{enumerate}

These metrics provide the basis of a quantifiable comparison of the
strength of similarities with statistical models and algorithms. Common
examples of such algorithms include random forests
\citep{hofmannBulletxtrctrAutomaticMatching2022} and congruent matching
profile segments
\citep{chenFiredBulletSignature2019, juOpenSourceImplementationCMPS2022}.
A large number of these algorithms are based on the maximized
cross-correlation function between pairs of signals. This is the metric
which we will use in this paper. However, this is not an actual
restriction, any other similarity metric would work similarly well, with
its usability only restricted by the metric's diagnostic ability.

Assume that \(X = \left\{X_t\right\}_{1 \le t \le N_X}\) and
\(Y = \left\{X_s\right\}_{1 \le s \le N_Y}\) are the observed surface
measurements (signals) of two land engraved areas (with \(N_X, N_Y\) the
number of the respective observations). The correlation between \(X\)
and \(Y\) is defined as the ratio of their covariance scaled by their
respective variances: \[
\text{corr} (X, Y) = \frac{\text{Cov(X, Y)}}{\sqrt{\text{Var}(X) \text{Var}(Y)}}
\] \(Y^{(k)}\) defines the \(k\)th lag of \(Y\) with
\(Y^{(k)}_s = Y_{k+s}\) with \(k \in [-M, M]\) and \(0 \le M < N_Y\).
The choice of \(M\) depends on the minimal number of values \(N_Y-M\)
used as a basis for an evaluation of the similarity of the two signals.
With that we define the maximized cross-correlation function
\(CCF_{\text{max}} (X, Y)\) as \[
CCF_{\text{max}} (X, Y) = \text{arg} \text{max}_{k \in [-M, M]} \text{corr}(X, Y^{(k)}).
\] Here, we use \(M=500\) for the alignment of signals. This corresponds
to a horizontal shift of \(\pm 500\) values (equal to
\(\pm 500 \times 0.645 \mu = .323 mm\)) corresponding to about a quarter
of a scan's width.

\begin{figure}

\centering{

\includegraphics[width=0.27\textwidth,height=\textheight]{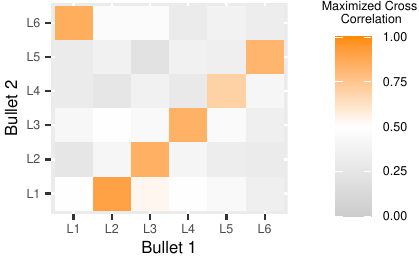}

}

\caption{\label{fig-land-matrix}Tile plot of all pairwise comparisons of
LEA signals from two different bullets.}

\end{figure}

When assessing the similarity of one bullet to another, a common
approach is to assemble all scores from comparing pairs of lands in form
of a square matrix and visualize it in form of a tile plot, see
Figure~\ref{fig-land-matrix}. The fill color encodes the score (here,
the ccf) between a pair of LEAs. Higher values indicate higher
similarity, shown in shades of orange. Tiles filled with grey values
indicate less similarity. The two bullets shown in the example are known
to have been fired through the same barrel. In this case, we expect six
pairs of lands with high similarities (in-phase), while all other pairs
(out-of phase) should result in low scores. This is exactly the pattern
that can be seen in Figure~\ref{fig-land-matrix}.

Numerically, we can summarise the matrix of pairwise LEA comparisons
into a single statistic by calculating averages of selected comparisons.
Here, we are using the (background-adjusted) maximum correlation score
\citep{juOpenSourceImplementationCMPS2022} between bullets \(B_1\) and
\(B_2\), given as the difference between the in-phase average and the
out-of-phase average: \[
\mathrm{\overline{CCF}_{diff}}(B_1, B_2) = \underbrace{\left[  \frac{1}{n} \sum_{(i,j) \in \mathcal{P}} c_{ij}\right]}_{\text{in-phase average}} - \underbrace{\left[  \frac{1}{n(n-1)} \sum_{(i,j) \notin \mathcal{P}} c_{ij}\right]}_{\text{out-of-phase average}},
\] where \(c_{ij}\) is the score between land \(i\) on bullet 1 and land
\(j\) on bullet 2, with \(1 \le i, j \le n=6\), where \({\cal P}\)
denotes the pairs of lands that capture the best alignment between
bullets \(B_1\) and \(B_2\).

A pairwise comparison of \(K\) number of bullets results in a set of
scores of size \(^KC_2 = \frac{1}{2}K(K-1)\) or \(^KC_2 + K\), if we
also consider to allow a comparison of a bullet to itself (done to
achieve an empirical assessment of the range of scores we can expect to
see for a particular type of ammunition and firearm). Different types of
visualizations of these set of scores are discussed in the next section.

\section{Visualization Framework}\label{sec-visuals}

\begin{figure}

\centering{

\includegraphics{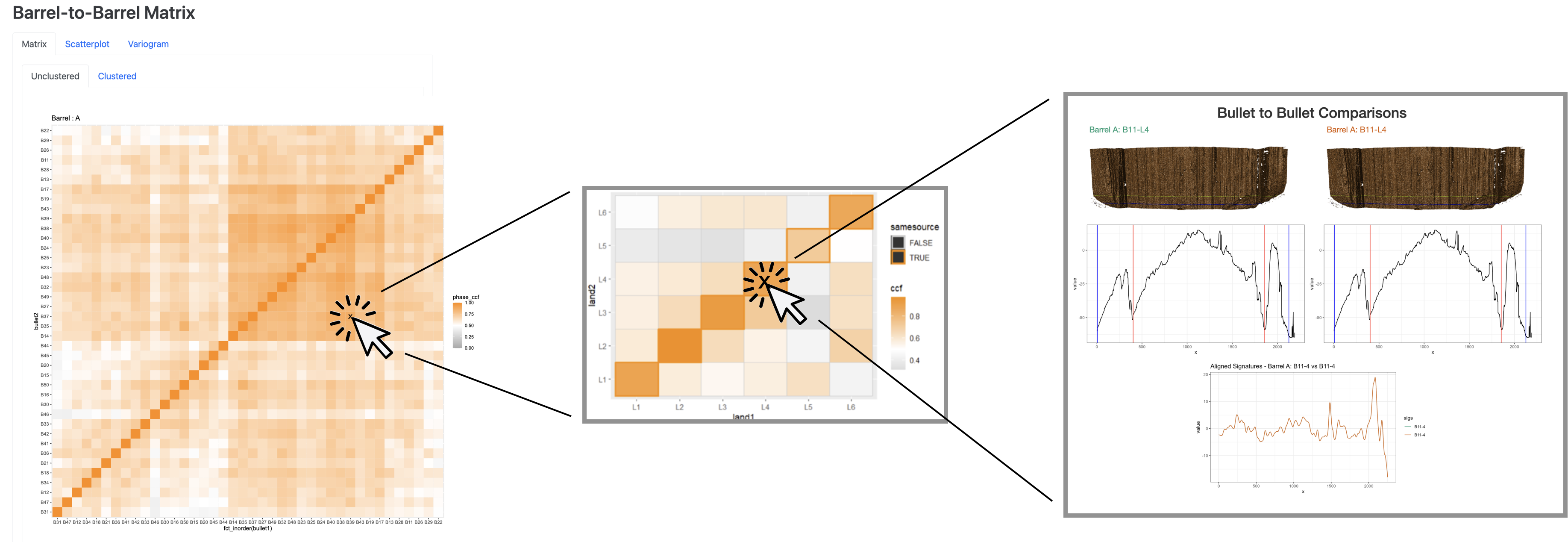}

}

\caption{\label{fig-tool-pipeline}Three connected levels of information.
From left to right, there is a tile plot of scores from all bullets in
one barrel, a tile plot of scores at the land-level for one pair of
bullets, and a set of diagnostic plots for comparing a single pair of
lands.}

\end{figure}%

As seen in the previous section, there are similarity scores at the
bullet-to-bullet level, there are scores at the land-to-land level, and
there are important diagnostics for individual pairs of lands. For any
given comparison, we have pertinent information at each of these levels
(see Figure~\ref{fig-tool-pipeline}). The statistical perspective
focuses on scores within the distribution of other, comparable scores,
while the focus in a forensic examination is on the individual. The idea
of this visualization tool is to connect these different levels and
perspectives for a seamless exploration. The forensic bullet comparison
visualizer (FBCV) is created in HTML using a combination of Javascript
and R code. This allows us to leverage the everyday familiarity of links
for implementing connections across levels of information. An
implementation of the FBCV showing all comparisons involving bullets
from barrel A can be found at https://tinyurl.com/y53n3mkm.

\begin{figure}

\centering{

\includegraphics[width=3.14in,height=\textheight]{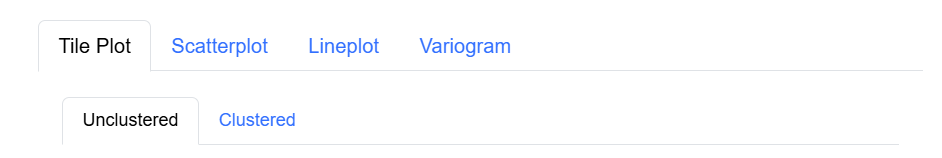}

}

\caption{\label{fig-framework-interface}Tabs in the Interface framework}

\end{figure}%

\begin{figure}

\centering{

\includegraphics{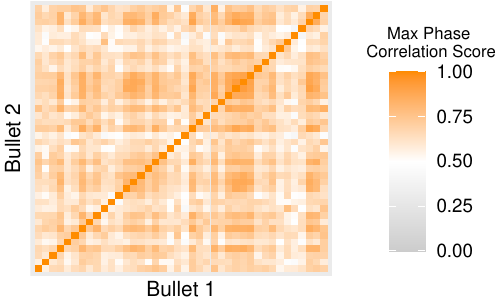}

}

\caption{\label{fig-barrel-matrix}Tile plot of the scores of all
pairwise comparisons of the 40 sequential bullets fired from barrel A}

\end{figure}

The main interface of the FBCV consists of a set of tabs with choices
for the visualization of the set of bullet scores, see
Figure~\ref{fig-framework-interface}, discussed in more detail next.
\hfill\newline \textbf{Tile plots} are our default choice for
visualizing all bullet comparisons. Figure~\ref{fig-barrel-matrix} shows
the 40x40 matrix of all pairwise bullet comparisons in for barrel A and
same-bullet scores on the diagonal. Each row and column corresponds to
the bullet involved in the comparison, each cell represents the maximum
phase correlation score for the respective comparison.

Note that this matrix is not static: the spatial area of each tile maps
interactively to the corresponding 6x6 land-to-land tile plot, such as
the one in Figure~\ref{fig-land-matrix}. When a user clicks on a square
in the 40x40 matrix, the FBCV retrieves the corresponding 6x6 matrix of
land-to-land scores, giving users more details on the land-to-land
comparisons that contribute to the score of the selected square.

The interactivity extends beyond the 6x6 matrix. By clicking on an
individual square within this matrix, additional information is provided
for the two LEAs and their comparison: renderings of the two LEA scans
with marked crosscut locations, plots of their profiles, and the aligned
signals. These web links directly map the different levels of
comparisons as shown in Figure~\ref{fig-tool-pipeline} to individual
comparisons and allow the user to move naturally between abstraction
levels.

\begin{figure}

\centering{

\includegraphics[width=0.4\textwidth,height=\textheight]{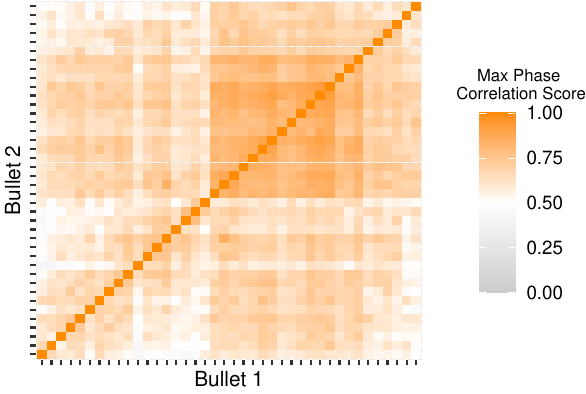}

}

\caption{\label{fig-ClusteredMatrix}Re-ordered tile plot for barrel A
scores.}

\end{figure}

\begin{figure}

\centering{

\includegraphics{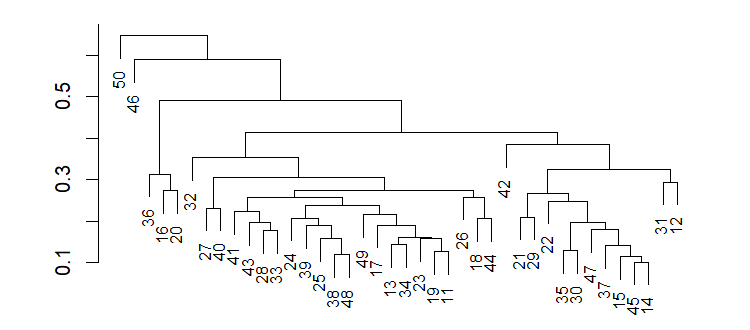}

}

\caption{\label{fig-dendrogram}The dendrogram for hierarchical
clustering}

\end{figure}

Ordering rows and columns in tile plots has a large impact on the
visualization. By clicking the \emph{Clustered} subtab, the user can
view an altered version of the original 40x40 tile plot. The ordering of
the rows and columns is based on a complete-linkage hierarchical
clustering of the score matrix. In Figure~\ref{fig-ClusteredMatrix} we
see two fairly distinct clusters. This version of the tile plot groups
bullets by their similarity, which helps to identify any significant
performance discrepancies in the data. The order of the bullets in
Figure~\ref{fig-ClusteredMatrix} follows the order of the dendrogram in
Figure~\ref{fig-dendrogram}. Also note that the interactivity of the
clustered tile plot is the same as for the original plot.

\hfill\newline

\begin{figure}

\centering{

\includegraphics[width=0.4\textwidth,height=\textheight]{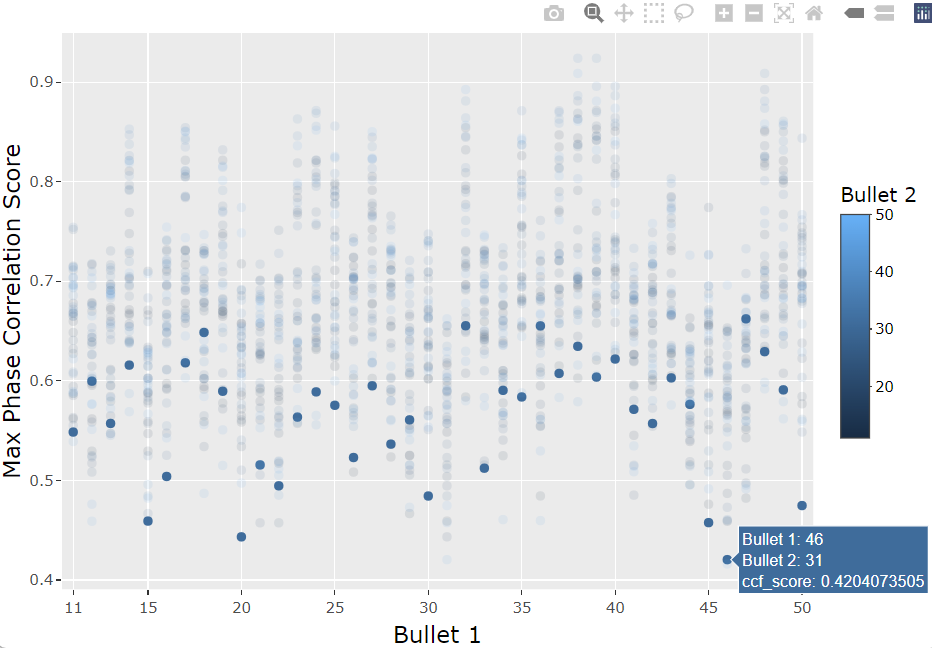}

}

\caption{\label{fig-scatterplot}An interactive scatterplot for bullet
comparisons}

\end{figure}

\textbf{Scatterplots} provide an alternative representation of the
scores: Figure~\ref{fig-scatterplot} shows an example of the default
scatterplot. The first bullet in the comparison is represented on the
x-axis, while the associated maximum phase correlation score is
displayed on the y-axis. Additionally, we use color to represent the
shot number of the second bullet. When the user hovers over a point, all
other points containing the second bullet in the comparison are
highlighted. This enables the user to identify bullets with poor scores
across the dataset or those exhibiting similar patterns across all
comparisons.

Clicking on a point again brings up the 6x6 matrix of land-to-land
comparisons resulting in the selected point's score.

\begin{figure}

\centering{

\includegraphics{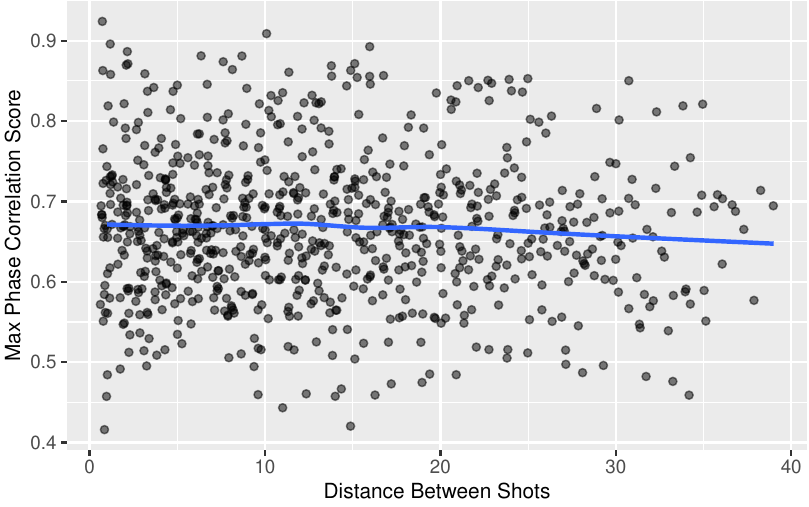}

}

\caption{\label{fig-variogramA}The variogram included in the
visualization framework}

\end{figure}

Clicking on the \emph{lineplot} tab brings up the same scatterplot, with
the key distinction that points representing the same second bullet in
the comparison are connected by a line. This visualization helps to
emphasize the relationships and trends between the points that share the
same bullet. \hfill\newline \textbf{Variograms} are used to represent
values as a function of the distance. In this context, the variogram
illustrates how the similarity of bullets is affected by the number of
bullets fired between them, shown in Figure~\ref{fig-variogramA}. The
x-axis represents the numerical distance between shots (11 vs.~12
corresponds to a distance of 1, while 11 vs.~50 is a distance of 39).
The y-axis represents the algorithmic score between the bullets. The
blue line shows a loess fit to capture the main trend. Clicking on any
point within the variogram leads to the same interactive pipeline as the
scatterplots and other visualizations.

These visualizations are integrated into a single HTML webpage,
providing a comprehensive view of the data and offering accessible
insights into scores.

\section{Use Cases}\label{use-cases}

\subsection{The Case of Barrel D}\label{the-case-of-barrel-d}

\begin{figure*}

\begin{minipage}{0.23\linewidth}

\centering{

\includegraphics{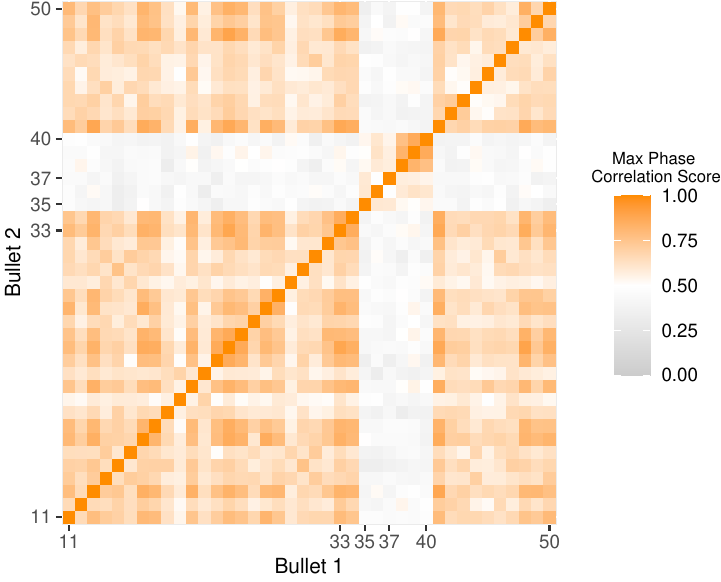}

}

\subcaption{\label{fig-matrixD-1}40x40 comparison matrix for firearm D}

\end{minipage}%
\begin{minipage}{0.03\linewidth}
~\end{minipage}%
\begin{minipage}{0.23\linewidth}

\centering{

\includegraphics{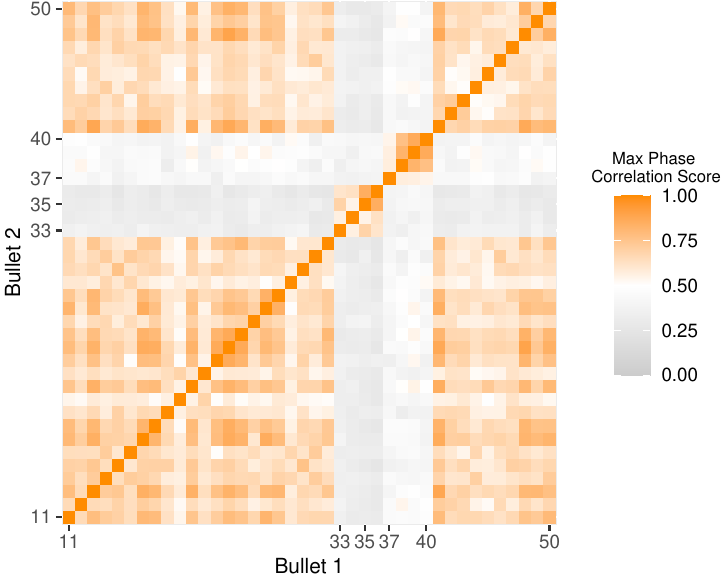}

}

\subcaption{\label{fig-matrixD-2}Replacing bullets 33-36 of barrel D
with scans of groove-engraved areas}

\end{minipage}%
\begin{minipage}{0.03\linewidth}
~\end{minipage}%
\begin{minipage}{0.23\linewidth}

\centering{

\includegraphics{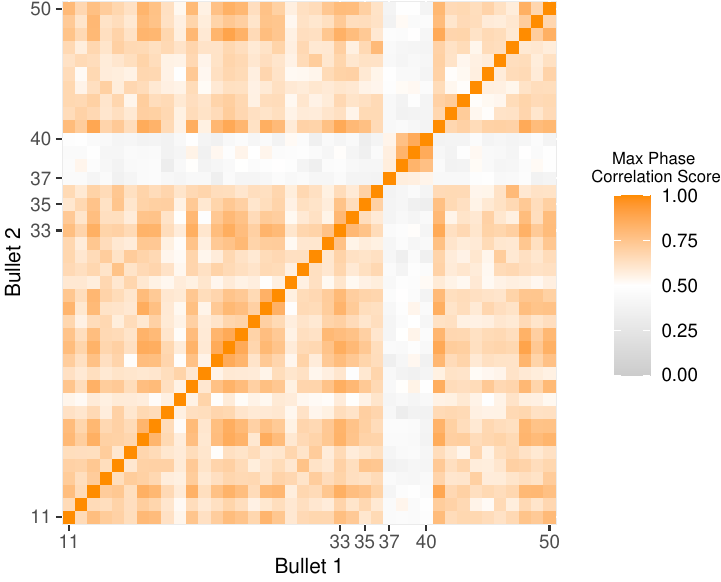}

}

\subcaption{\label{fig-matrixD-3}Replacing bullets 33-36 of barrel D
with LEAs rescans}

\end{minipage}%
\begin{minipage}{0.03\linewidth}
~\end{minipage}%
\begin{minipage}{0.23\linewidth}

\centering{

\includegraphics{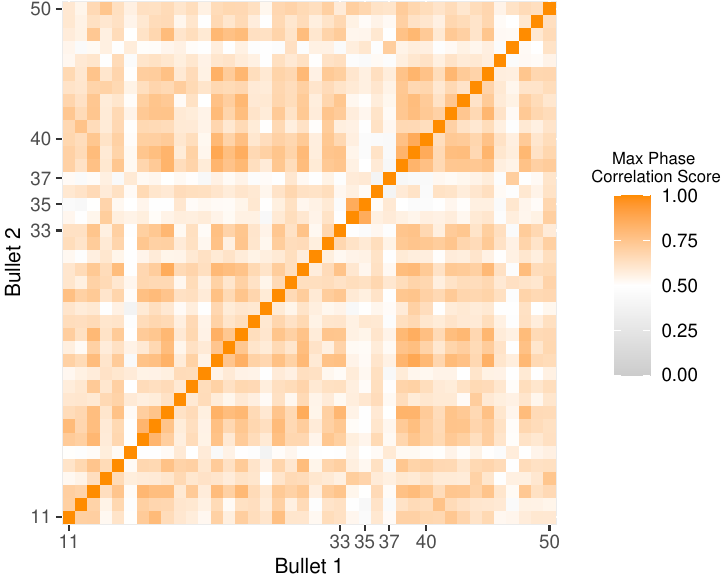}

}

\subcaption{\label{fig-matrixD-4}Replacing bullets 35-50 of barrel C
with original scans from barrel D}

\end{minipage}%

\caption{\label{fig-matrixD}Overview of the adjustments made throughout
the data cleaning process and their impact on the matrices}

\end{figure*}

The provided visualizations have shown scores for barrel A, but not all
firearms displayed such straight-forward results. One such case was the
scores for barrel D. After running the pairwise comparisons and
analyzing the visualization framework, it became apparent that there was
an error in the analysis for bullets 35-40. These bullets performed well
when compared to each other but did not score highly compared to the
other bullets shot from this firearm, as shown in
Figure~\ref{fig-matrixD-1}.

\vspace{-10pt}

\hfill\newline\noindent \textbf{Comparing Groove-Engraved
Areas}\hfill\newline One potential reason for this suboptimal
performance may stem from the groove-engraved areas (GEAs) being scanned
rather than the land-engraved areas. Note that these areas are usually
not used for examinations because grooves preserve marks from the tool
they are made. When firearms are manufactured, a broaching tool is used
to create the grooves for the rifling. This incorporates marks specific
to the tool on the surface of the barrel. Because the same broaching
tool is used for multiple barrels, marks on grooves are not specific to
the firearm, limiting the ability to conclusively link striations on
groove-engraved areas of a bullet to a particular firearm.

\begin{figure}

\begin{minipage}{0.50\linewidth}

\centering{

\includegraphics{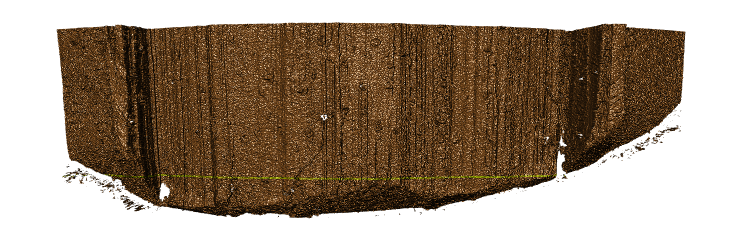}

}

\subcaption{\label{fig-LEA35}Original land from bullet 35}

\end{minipage}%
\begin{minipage}{0.50\linewidth}

\centering{

\includegraphics{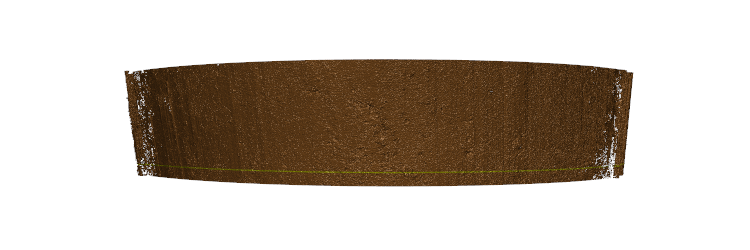}

}

\subcaption{\label{fig-GEA}Scan of a groove from bullet 35}

\end{minipage}%

\caption{\label{fig-bulletScans}Comparing the land-engraved area and the
groove-engraved area of bullet 35 from barrel D.}

\end{figure}%

When analyzing the scans of groove-engraved areas (GEAs), it became
apparent that the usage of grooves was not the root of the problem.
Figure~\ref{fig-LEA35} below shows the original scan from Bullet 35,
while Figure~\ref{fig-GEA} shows a rescanned groove for that same
firearm. Notably, the scan from GEAs has fewer topographical protrusions
than the original image, indicating a smoother surface profile.

We then conducted an analysis where bullets 33-36 in the original data
were replaced with comparisons using the grooves. For this test, bullets
33 and 34 act as a control - they represent scans that were already
producing expected results in the dataset. Bullets 35 and 36 represent
two bullets performing poorly in barrel D. The 40x40 matrix was
recreated using our interactive framework, shown in
Figure~\ref{fig-matrixD-2}.

This visualization reinforces the conclusion that the discrepancy in the
max phase correlation score was not due to grooves being scanned instead
of lands. Not only are bullets 35 and 36 performing worse than before,
but the scores on our control bullets - 33 and 34, also dropped
significantly. Thus, we can conclude that the inaccuracy of the original
data is not because groove areas were analyzed instead of lands.
\hfill\newline\noindent \textbf{Rescanning Land-Engraved
Areas}\hfill\newline Our following action was to rescan and process the
3D topographical imaging on the bullets in question. The same control
and test groups were used as the groove comparisons. Bullets 33 - 36
were rescanned, with 33 and 34 as the control. Replacing those
comparisons in the interactive framework showed a drastic difference in
results. In the 40x40 matrix, the max phase correlation scores of the
rescanned bullets aligned closely with those the other bullets fired
from firearm D, as shown in Figure~\ref{fig-matrixD-3}. Furthermore, no
significant change was found in the scores of the control group. When
utilizing other parts of \thetool, such as the 6x6 matrix and looking at
the raw scans, these comparative results were reinforced. The alignment
of signals for the rescans showed significant improvement compared to
the previous alignment among barrel D.

Thus, the observed discrepancy in performance stems not from the
algorithmic process itself but from inconsistencies in the raw scans
utilized in the initial data processing. This could be due to various
factors, including mislabeling or inadequate scanning. However, if this
is a case of mislabeling scans, it raises questions regarding the
provenance of the original scans. \hfill\newline\noindent
\textbf{Closing the Loop}\hfill\newline To answer whether the data was
mislabeled, we compared the original scans of bullet D to data obtained
from all of the 12 other firearms in the Houston dataset. We selected
one bullet from each firearm that performed exceptionally compared to
the other bullets from that firearm (12 bullets in total). Those bullets
were then compared to both each other and bullet 39 from firearm D, one
of the originally poor-performing bullets.

\begin{figure}

\centering{

\includegraphics{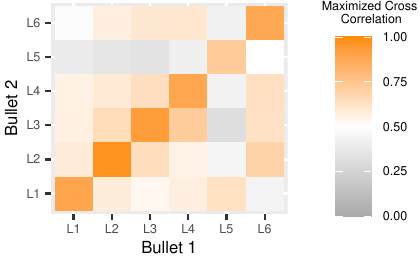}

}

\caption{\label{fig-CD-Comparison}Replacing 6x6 matrix from bullet C
with bullet D}

\end{figure}

We found that 11 of the 12 bullets showed poor performance when compared
to the selected bullet from firearm D. The exception was the bullet
selected from firearm C. When comparing that bullet to bullet 39 of
barrel D, the algorithmic results showed that the two bullets were
likely fired from the same weapon. Figure~\ref{fig-CD-Comparison} shows
the 6x6 matrix from our framework tool when comparing the selected
bullet from firearm C to bullet 39 of barrel D. The alignment between
these two bullets is extremely strong, even stronger than many other
bullet comparisons where both bullets originated from firearm C.

It is also important to note that the selected bullets did not perform
well when compared with each other (i.e., results from the chosen bullet
from firearm A did not match the bullet from firearm B). This implies
the similarity between the selected bullet from firearm C and the
poor-performing bullet from firearm D cannot be attributed to
well-performing bullets being accurate among comparisons with other
weapons. Thus, we have strong evidence that the bullets were mislabeled,
with the poorly performing bullets originating from barrel C.

This notion is strengthened by our visualization tool. When the
comparisons for bullets 35-40 are substituted with those initially
labeled as bullets 35-40 fired by barrel D, there is no significant
deviation in performance according to the 40x40 matrix shown in
Figure~\ref{fig-matrixD-4}. While there may appear to be a dip in
performance around this area, the differentiation happens in bullets
34-37, which is not the complete scope of the substituted bullets.
Overall, these scans perform very well compared to other bullets shot by
firearm C. Thus, we can conclude that bullets 35-40 were mislabeled and
shot by firearm C, not D.

\subsection{Bullet Distance Analysis}\label{bullet-distance-analysis}

\begin{figure}

\centering{

\includegraphics{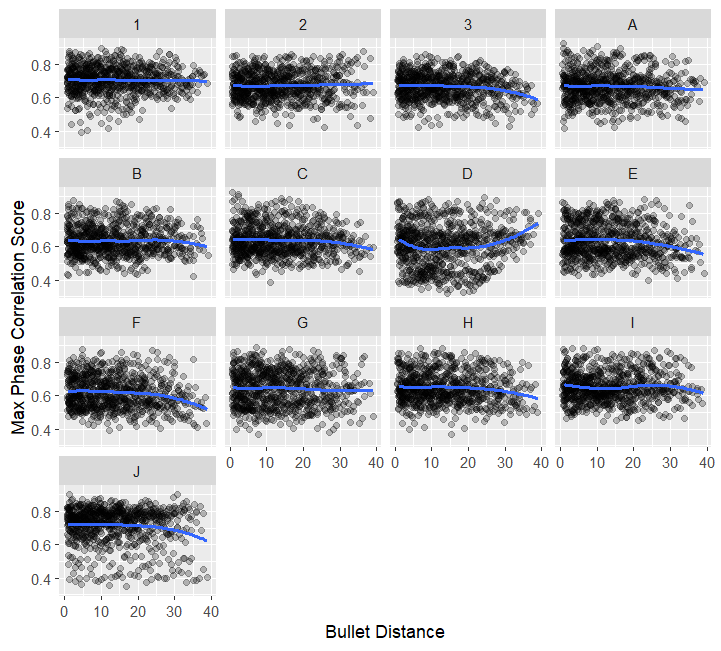}

}

\caption{\label{fig-variograms}Variograms across all firearms}

\end{figure}

A key focus of this tool was evaluating whether model performance
changes as more bullets are fired from a weapon. Using variograms, we
provided a visual representation of this performance.
Figure~\ref{fig-variograms} illustrates variograms for all firearms in
the Houston dataset.

We hypothesized that model performance would decline as the distance
between shots fired increases. This does appear to be the case in some
firearms, specifically firearms 3, B, C, E, F, and H. However, other
barrels, such as 1, 2, A, and G show little to no deviation in
performance as distance increases.

Furthermore, some firearms show odd behaviors. Firearm D shows an upward
trend, although this may be attributed to the aforementioned mislabeling
issue. Firearm I exhibits an unusual pattern, where the scores initially
increases with distance, followed by a downward trend at greater
distances. Finally, firearm J shows a high frequency of comparisons with
markedly low performance relative to the other bullets in the analysis.
Using the visualization framework, we recognized that this behavior can
be attributed to markedly low scores using comparisons from two bullets
- bullets 46 and 50. Analysis similar to the previous case in barrel D
is needed to ascertain the cause of such low scores. Overall, given the
varied results of this data, we cannot make any definitive claims
regarding model performance across bullet distance.

\section{Limitations}\label{limitations}

This study faces a number of limitations, chiefly regarding file
storage. Because of the large number of comparisons, a considerable
amount of files were rendered for each stage in the FBCV's process. Each
firearm contains 1600 bullet-to-bullet comparisons. Then, for each
bullet-to-bullet comparison, there are 36 sub-comparisons made at the
land level, corresponding to the six lands of each bullet. Therefore, we
must process \(1600 \times 36 = 57,600\) png images at the land level.

At the lowest level of the FBCV, an image is processed for each pf the
lands being compared, the cutoffs of the LEA for both lands, and the
aligned signals. Therefore, 288,000 background PNG files are processed
before they are aggregated into HTML format. To mitigate the storage
burden, we employed a strategy to avoid creating new background files
for each HTML rendering. For the 40x40 matrix, the 6x6 land-to-land
matrix, and the bullet comparison informational HTML, we used a single
background file for each type of HTML. This background file was reused
across different comparisons, enabling more efficient rendering of the
FBCV without duplicating files for each individual bullet comparison.

Despite these adjustments, the substantial number of files presents a
significant storage challenge, and Github was unable to accommodate more
than the number of files associated with one firearm. Consequently, the
figures presented in this analysis primarily focus on firearm A.
Furthermore, the need to render such a large number of files complicates
the process of making rapid updates to the scans, as any modification
typically requires re-rendering a large number of figures. Still, as
demonstrated in previous examples, the framework does allow for updates
as changes are introduced.

\section{Conclusion}\label{conclusion}

This paper presented an interactive framework for analyzing algorithmic
comparisons of whether two bullets were fired from the same firearm. The
framework includes various visualizations that allow the user to assess
algorithmic performance at a broader scope while also diagnosing issues
at every level of the comparative analysis process. The FBCV was used to
analyze algorithmic performance on the Houston dataset. It successfully
identified a problematic error in the comparison process, and
investigative steps were taken to discern the cause of the error, which
is attributed to mislabeling. In the future, this visualization
framework can provide summary overviews of algorithmic performance and
diagnose problems in the data processing of forensic bullet analysis. By
offering an interface that is intuitive and accessible, it presents an
option that can support forensic examiners and lead to more accurate
forensic analysis.

  \bibliography{bibliography.bib}

\end{document}